\documentclass[prl,twocolumn,showpacs,nofootinbib]{revtex4}
\usepackage[dvips]{graphicx}
\usepackage{color,array,dcolumn}
\usepackage{amsmath}
\usepackage{amssymb}

\begin{document}

\author{P.A. Marchetti}
\affiliation{Dipartimento di Fisica ``G. Galilei'', INFN, I-35131 Padova, Italy}

\author{Z.B. Su}
\author{ L. Yu}
\affiliation{Institute of Theoretical Physics and Interdisciplinary Center of Theoretical
Studies,\\
Chinese Academy of Sciences, 100080 Beijing, China}
\title{SPIN-CHARGE GAUGE SYMMETRY: A WAY TO TACKLE HTS CUPRATES?}
\date{\today }

\begin{abstract}
We propose an explanation of several experimental features of
transport phenomena in the normal state of high $T_c$  cuprates in
terms of a spin-charge gauge theory of the 2D $t$-$J$ model. The calculated
   doping $(\delta)$-temperature ($T$) dependence for a number of physical
   quantities is found in  qualitative agreement with data. In particular, we recover: in the ``pseudogap phase''  the metal-insulator
    crossover of the in-plane resistivity and of the NMR ``relaxation time'' $(T_1T)^{63}$
     and the insulating behavior of the out-of-plane resistivity;
in the ``strange metal phase''   (at higher $T$ or $\delta$)  the linear in $T$ behavior
     of the above quantities;
     the appearance of maxima in the  in-plane far-infrared conductivity in strongly
      underdoped and overdoped samples.
\end{abstract}

\maketitle
\section{The spin-charge gauge approach}
 We assume as a model for CuO layers in high $T_c$ cuprates the 2D $t$-$J$ model.  However, neglecting
from beginning the n.n.n. hopping all features depending on the
detailed structure of the Fermi surface are clearly lost. Gauging  with a Chern-Simons gauge field action the spin and the charge global symmetries of the $t$-$J$ model one obtains a gauged $t$-$J$ model strictly equivalent to the original \cite{Fro}. The spin-charge gauged model suggests an improved ``Mean Field
Approximation '' (MFA) based on a formal spin-charge decomposition of the electron field \cite{Mar}. In this approach, the low energy effective action describes spin $1/2$ gapful bosonic spinons,
with a non-linear sigma (NL$\sigma$) model action, and  charged fermionic holons interacting via a slave-particle gauge field.
Peculiar to our approach and present both in the ``pseudogap'' and ``strange metal phases'' (although here less effective) is the spinon gap $m_s\sim J(\delta |\ln \delta|)^{1/2}$. It is due to spin vortices attached by the spin gauge field in MFA to the empty-site positions and at high temperature it receives the standard  thermal correction of the NL$\sigma$ model. The theoretically derived $\delta-T$ dependence of the spinon gap up to $\sim 500 K$ is in qualitative agreement with the behavior of the AF correlation length in underdoped LSCO \cite{belk}.

 In the ``pseudogap phase'' (PG, at  low $\delta$ and $T$) the holons have ``small'' Fermi surface $(\epsilon_F \sim  t \delta)$
 and a ``Dirac structure'' induced via Hofstadter mechanism by the $\pi$-flux lattice of the charge-gauge field in MFA, with bare dispersion $$\epsilon^{PG}(\vec p)=\pm 2t \sqrt{{\cos^2(p_x)}+{\cos^2(p_y)}}$$
 restricted to
the magnetic Brillouin zone.
The main  effect of gauge fluctuations is to induce a  a shift in the mass of spinons \cite{Dai} 
$
m_s \rightarrow M=(m_s^2- i c {T / \chi})^{1/2}
\nonumber
$,
where $c$ is a constant and $\chi$ the diamagnetic
susceptibility, thus introducing a
dissipation $\sim T$ . The competition between the two energy scales,
$m^2_s$ and $T/\chi$, is the root in our approach of many crossover  phenomena
peculiar to in-plane transport properties of the  ``pseudogap phase'' \cite{mdosy}. In particular it yields an explanation of the metal-insulator crossover (MIC) observed in underdoped, non-superconducting cuprates as temperature decreases \cite{Taka}, \cite{wuyts} and a similar crossover in superconducting cuprates when a strong magnetic field suppresses superconductivity \cite{Fou}. Hence, in our approach the MIC is due to correlation effects, not to a disorder-induced localization which would be hard to reconcile with the experimental fact that $k_F \ell$ at MIC ranges from O(0.1) to O(10), i.e. well below and well above Ioffe-Regel limit. 
Furthermore the gauge interaction induces the
binding of spinon and anti-spinon into a massive damped magnon resonance and (close to the Fermi surface) of spinon and holon into an ``electron'' resonance with a $T^{1/6}$-dependent wave function renormalization constant $Z$ and inverse life-time $\Gamma \sim |M| \sin \arg(M)$. This
energy scale, i.e. the recombination rate,
dominates the out-of-plane resistivity, yielding an
insulating behavior. The ``Dirac dispersion'' of holons induces a reduction of the spectral weight  outside the magnetic Brillouin zone, reminiscent of the ``Fermi arcs'' found in ARPES in underdoped cuprates \cite{largeFS}, see Fig.1.

Around a temperature $T^*$ (identifiable as the inflection point of in-plane resistivity) the $\pi$-flux lattice of the charge  gauge field melts and we enter in the ``strange metal phase'' (SM, at high $\delta$ or $T$) . In such phase \cite{mosy}, holons have ``large'' Fermi surface $(\epsilon_F \sim t (1-\delta))$
 and standard dispersion. 
The gauge fluctuations induce still the formation of a strongly overdamped magnon resonance, with inverse life-time $\sim T Q_0/(\chi m_s^2) \sim T^{4/3}$ and an ``electron resonance'' with inverse life-time , $\Gamma$, of the same order and $Z \sim \sqrt{m_s Q_0}/J \sim T^{1/6}$. $Q_0^{-1} =(\chi/\kappa T)^{1/3}$ is the typical scale of gauge fluctuations, where $\kappa$ is their Landau damping, and can be thought of as a sort of anomalous skin depth.
Consistently with experimental data, linear in $T$ behaviour of in-plane and out-of-plane resistivity and  spin relaxation time $(TT_1)^{63}$ is recovered at large enough $T$, as a consequence of the above life-time and the ``effectiveness'' of the gauge fluctuations in a slab of momenta $Q_0 \sim T^{1/3}$ around the Fermi surface: in fact the conductivity derived from Boltzmann transport theory would be $\sigma_0 \sim \Gamma ^{-1}$, but due to ``effectiveness'' the physical conductivity is $\sigma \sim \sigma_0 Q_0 \sim T^{-4/3} T^{1/3} \sim T^{-1}$ .

\maketitle
\section{Behavior of physical quantities}

{\sl In-plane resistivity}

It is calculated via Ioffe-Larkin addition rule \cite{Iof}, a typical feature of slave-particle gauge theories: $\rho=\rho_s + \rho_h$,
where $\rho_s$ is the resistivity of the spinon-gauge subsystem and $\rho_h$ of the holon-gauge subsystem, subdominant.  Kubo formula applied to $\rho_s$ gives in the PG in the range $ m_s Q_0\lesssim T/\chi \lesssim m_s^2  $ (roughly from few tenth to few hundreds K) where the approximations made appear reliable,

\begin{equation} 
  \rho \sim \rho_s\sim {|M|^{1/2} \over \sin(\arg{M \over 2})} \sim
   \left\{
   \begin{array}{ll}\label{rpg}
      T^{-1}
      & m_s^2 >> {c T \over \chi}  \\ 
      \sim T^{1 \over 4}
      &m_s^2 \sim {c T \over \chi}
   \end{array}
   \right  .
\end{equation}
Thus one recovers the metal-insulator-crossover (MIC) and an
inflection point $T^* \sim \chi m_s^2 \sim |\ln \delta|$ (since $\chi \sim \delta^{-1}$in PG) at higher temperature, found
also experimentally \cite{Taka}; see Fig. 2. One easily verifies that the
normalized resistivity
$\rho_n=(\rho-\rho(T_{MIC}))/(\rho(T^*)-\rho(T_{MIC}))$ is a function
only of the ratio $T/(\chi m_s^2)$. Therefore a prediction of the theory is a universal
behaviour which in fact can be verified in experimental data and was already empirically remarked in \cite{wuyts}, \cite{raffy}. See Fig. 3,4.
The introduction of a magnetic field perpendicular to the plane yields a shift of the MIC at higher temperature and this in turn produces a big positive transverse magnetoresistance \cite{magneto}, as found in experiments \cite{lacerda}.

In SM one obtains, for $m_s, \Gamma \lesssim Q_0$ (roughly from few tenth to 500 K), where the approximation made in our approach are reasonable:
\begin{equation}\label{asympt}
\rho_s\simeq 2 \lambda Q_0^{-1} (\Gamma +  m_s^2/ \Gamma)
= \frac{T}{\chi m_s^2} \lambda^2 + \frac{32 m_s^4 \chi}{T Q_0^2},
\end{equation}
where $\lambda \sim O(1)$ is a constant.
In the high temperature limit $Q_0 \gg m_s$, the damping rate in
(\ref{asympt}) dominates over the spin gap $2m_s$ and the spinon
contribution to resistivity is linear in $T$, with a slope $\alpha
\simeq (1-\delta)/(\delta |\ln
\delta|)$, see Fig. 5.
 Lowering the temperature, the second term in
(\ref{asympt}) gives rise first to a superlinear behavior and
then,  at the margin of validity of our approach, an
unphysical upturn. The deviation from linearity is due to the spin
gap effects and is cutoff in the underdoped samples by the
crossover to the PG phase. We expect that physically in the
overdoped samples it is cutoff  by a crossover to a FL``phase''.
 The temperature dependence of the spinon mass yields
  a bending at high temperature, stronger for lower dopings, as visible in the resistivity data
   at constant volume for LSCO \cite{sundqvist}; this effect is masked in the resistivity data at
    constant pressure \cite{Taka} by thermal expansion.

{\sl In-plane IR electronic AC conductivity}

We compute the electronic
AC conductivity via Kubo formula, in the two limits $\Omega << T$, $T <<
  \Omega$, where $\Omega$ is the external frequency. It turns out that up to the logarithmic accuracy one can pass from the first
  to the second limit by replacing  $T$ with $\Omega$ in $Q_0$ and $\Gamma$ ( we denote the
  obtained quantities by $Q_\Omega, \Gamma_\Omega$) and rescaling $\Gamma$ by a positive multiplicative factor
  $\tilde\lambda \lesssim 1/2$.
In PG, in the range $T \lesssim \Omega \lesssim J m_s^2$ the AC conductivity 
can be approximately obtained from $\rho^{-1}$ derived from eq. (\ref{rpg}) substituting $T$ with $\tilde\lambda \Omega$. Hence it exhibits a broad maximum at a frequency , $\Omega_{MIC}$, corresponding to a temperature slightly higher than $T_{MIC}$, as in fact experimentally seen in \cite{basov}, see Fig. 6, 7. An $a-b$ anisotropy of $T_{MIC}$ and $\Omega_{MIC}$ found in underdoped LSCO \cite{basov} would be generated naturally in the above scheme by an $a-b$ anisotropy of the magnetic correlation length \cite{anisotropy}. The anisotropy of $T_{MIC}$   strengthens our interpretation of correlation-induced MIC, since a disorder-induced localization is expected to have a unique MIC temperature. 
As $T$ becomes greater than $\Omega$ the AC conductivity becomes approximately $\Omega$ independent.  
 
In SM, for   $\Omega << T$ we have
\begin{equation}
\sigma(\Omega,T) \sim \frac{Q_0}{i (\Omega -2 m_s) +\Gamma} \sim \frac{1}{i (\Omega -2 m_s)
 T^{-1/3} + T},
\end{equation}
while for $T << \Omega$
\begin{equation}
\sigma(\Omega,T) \sim \frac{Q_\Omega}{i (\Omega -2 m_s)+ \tilde\lambda \Gamma_\Omega}
 \sim \frac{1}{i (\Omega -2 m_s) \Omega^{-1/3} + \Omega}.
\end{equation}
 From the above formulas the following features found experimentally in overdoped LSCO \cite{startseva} and BSCO \cite{lupi} 
 are easily derived: besides the standard tail $\sim \Omega^{-1}$, the effect of replacing  $Q_0$ by $Q_\Omega$
 is to asymmetrize the peak at $2 m_s$ appearing in Re$\sigma(\Omega,T)$ for $\Omega << T$
 and to shift it towards lower frequency, see Fig. 8.

{\sl Spin-lattice relaxation rate} $(1/T_1T)^{63}$

It is calculated via Kubo formula.
In the PG one finds:
\begin{eqnarray} \nonumber
 &({1 \over T_1T})^{63}  \sim (1-\delta)^2 |M|^{-\frac{1}{2}}(a\cos(\frac{\arg M}{2})+ \\ 
 & b\sin(\frac{\arg M}{2})) \sim
   \left\{
   \begin{array}{ll}
      T
      & m_s^2 >> {c T \over \chi}  \\ 
      \sim T^{-{1 \over 4}}
      &m_s^2 \sim {c T \over \chi}
   \end{array}
   \right .
\end{eqnarray}

Therefore we obtain a broad peak as observed in some cuprate \cite{Ber}, see Fig 9.
In SM one finds
\begin{equation}
\label{tt163}
(\frac{1}{T_1T})^{63} \sim (1-\delta)^2 \rho_s^{-1}.
\end{equation}

Therefore, in the high temperature limit we recover the linear in
$T$ behavior for
 $(T_1T)^{63}$ and at high doping / low temperature
 the superlinear deviation,  also found experimentally in overdoped samples of LSCO.\cite{berthier}, see Fig. 10.
 Furthermore, the factor $(1-\delta)^{-2}$ weakens the doping dependence of the slope as compared
 with the resistivity curves, in agreement with the experimental data. 

{\sl Out-of-plane resistivity}

We assume an incoherent transport along c-direction. $\rho_c$ thus is dominated
by virtual hopping between layers and it is
calculated via Kumar- Jayannavar formula \cite{Kumar}: 
\begin{equation} \label{kuma}
   \rho_{c} \sim \frac{1}{\nu}\left(\frac{1}
   {\Gamma}+\frac{\Gamma}{t_c^2 Z^2}\right),
\end{equation}
 where $\Gamma$ is the inverse electron life time, given here by the spinon-holon recombination rate, $t_c$ is an effective average hopping in the $c$-direction and $\nu$ the density of states at the Fermi energy.
In PG the first term in (\ref{kuma}) dominates, yielding the insulating behaviour:
\begin{equation} \nonumber
\rho_c\sim (|M| \sin(\arg M))^{-1}
  \sim
   \left\{
   \begin{array}{ll}
      T^{-1}
      & m_s^2 >> {c T \over \chi}  \\
      \sim T^{-{1 \over 2}}
      &m_s^2 \sim {c T \over \chi}
   \end{array}
   \right .
\nonumber \end{equation}.
This crossover reproduces a knee experimentally found in \cite{knee}, see Fig. 11. The very strong decrease with $T$ of the anisotropy ratio $\rho_c (T)/\rho_{ab} (T)$ in PG is also well reproduced by the above formulas, see Fig. 12.

In the SM phase the second, metallic, term in (\ref{kuma}) dominates and
substituting $\Gamma$ and $Z$ one recovers the $T-$linearity in the
``incoherent regime'' $\Gamma \gg t_c Z$:
\begin{equation}\label{rc}
\rho_c \simeq \frac{J^2}{t_c^2 m_s
\nu}\frac{T}{\chi m_s^2}\simeq
\frac{T}{(\delta|\ln\delta|)^{3/2}}. \end{equation}
The second term in eq. (\ref{kuma}) yields
possibly a minimum at low $T$ (if in the admissible range) see Fig. 13.
 Unfortunately, we don't have a reliable method to estimate the
(extrapolated) $T=0$ intercept of $\rho_c(T)$ which is large when
compared with the corresponding intercept for  $ab-$plane
resistivity,  so we cannot
extract safely the anisotropy ratio $\rho_c (T)/\rho_{ab} (T)$.
However, if the minimum in $\rho_c$ is at
higher temperature than the (unphysical) minimum in $\rho_{ab}$,
then one can derive the fast decrease of the anisotropy ratio at
low temperature found experimentally,\cite{nakamura} , see Fig. 14.

\vspace{1pc}

We would like to sincerely thank J.H. Dai, L. De Leo and G. Orso for their collaboration at different stages of this project.

%\begin{figure}
%\begin{center}
%\includegraphics[height=6cm,angle=0]{Marchett1a.eps}
%\caption{Calculated inverse magnetic correlation length versus temperature for% $\delta=0.01, 0.03,0.04$. {\sl Left}: Inverse magnetic correlation length for% LSCO samples from
%Ref.  \onlinecite{belk}.} \label{Fig. 1}
%\end{center}
%\end{figure}

\begin{figure}
\begin{center}
\includegraphics[width=6cm]{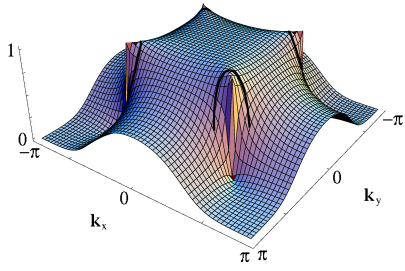}
\caption{Angle-dependent spectral weight of the electron propagator. The thick lines close to 
 ($\pm\pi/2,\pm\pi/2$) represent the region of FS with spectral weight larger than 1/2 for $\delta 
 \sim 0.05$}
\label{Fig. 2}
\end{center}
\end{figure}

\begin{figure}[tbp]
\includegraphics[width=6cm,angle=0]{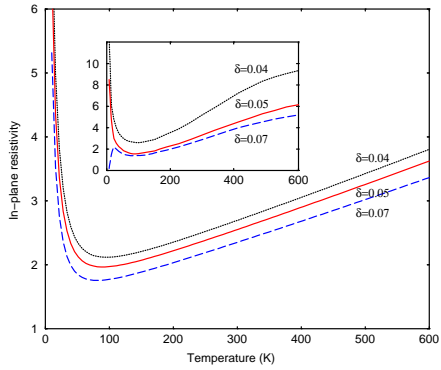}
\caption{ The calculated temperature dependence of in-plane resistivity for various 
dopings $\delta$ in comparison with the corresponding experimental data (inset) on La$_{2-\delta}$ Sr$_\delta$ Cu O$_4$
in units of $m\Omega cm$, taken from Ref. \onlinecite{Taka}.}
\label{Fig. 3}
\end{figure}

\begin{figure}
\begin{center}
 \includegraphics[width=6cm]{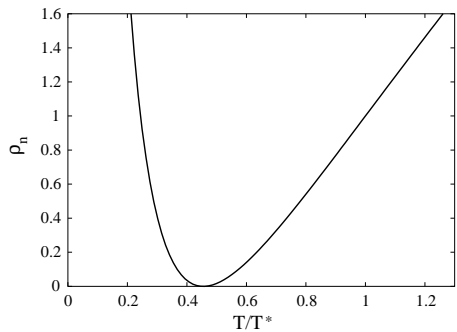}
 \caption{ Calculated ``normalised'' resistivity $\rho_{n}$ versus reduced temperature $T/T^{*}$ (see text for explanation).}
\label{Fig. 4}
\end{center}
\end{figure}

\begin{figure}[tpb]
%\begin{center}
 \includegraphics[width=6cm]{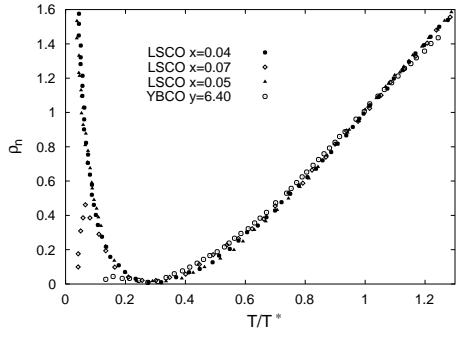}
 \caption{ Temperature dependence for $\rho_{n}$ in underdoped LSCO ( Extracted from
 Ref. \onlinecite{Taka}) and YBCO (Extracted from Ref.
 \onlinecite{wuyts})}
\label{Fig. 5}
%\end{center}
\end{figure}

\begin{figure}[tpb]
%\begin{center}
\includegraphics[height=6cm,angle=0]{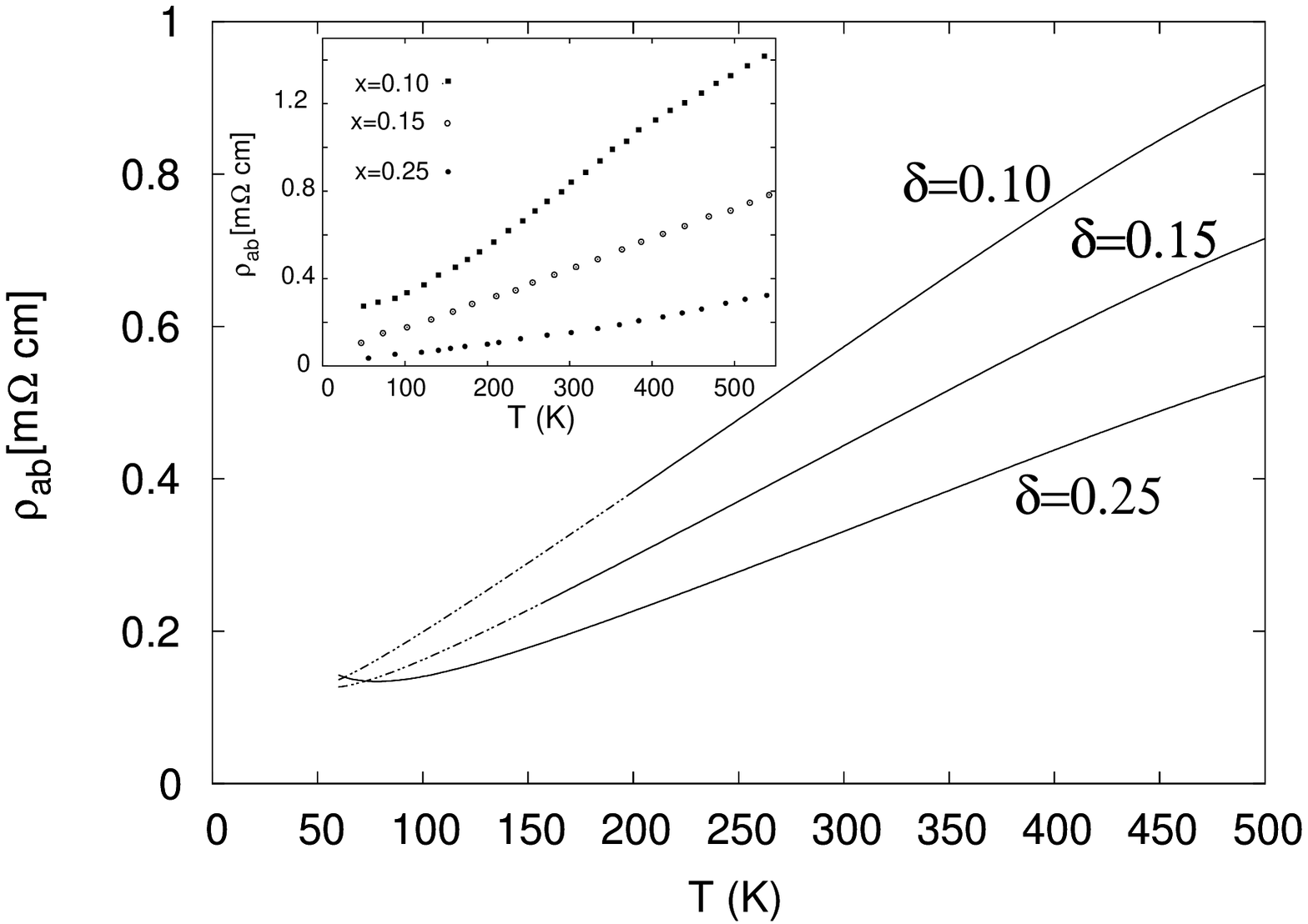}
\caption{Calculated in-plane resistivity as a function of
temperature for different hole concentrations.  Below the
pseudo-gap temperature $T^*$, the curve is shown in dashed line.
{\sl Inset:} In-plane resisitivity versus $T$ measured in LSCO
crystals with different Sr content $x$, taken from the work of
Takenaka {\it et al.},  Ref. \onlinecite{sundqvist}}.\label{Fig. 6}
%\end{center}
\end{figure}

\begin{figure}[tbp]
\includegraphics[width=4cm,angle=270]{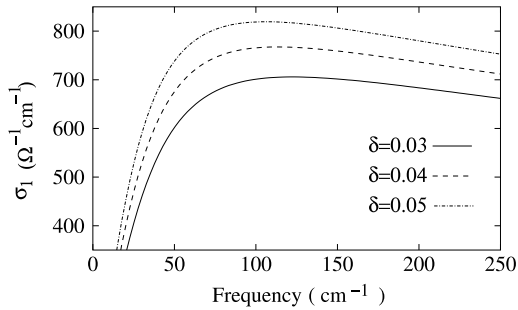}
\caption{Theoretically calculated frequency dependence of the AC
conductivity for different dopings: $\protect\delta=0.03$ (full line), $%
\protect\delta=0.04$ (dashed) and $\protect\delta=0.05$ (dotted).}
\label{Fig. 7}
\end{figure}

\begin{figure}[tbp]
\includegraphics[width=6cm,angle=270]{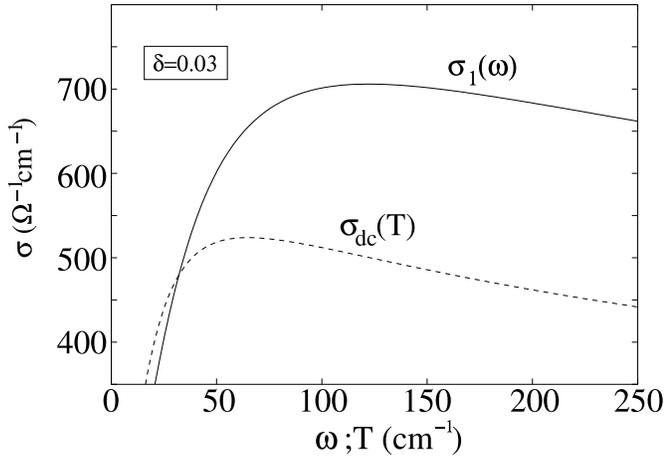}
\caption{Calculated frequency dependence of the AC conductivity for $\protect%
\delta=0.03$. Also shown is the corresponding DC conductivity as a function
of temperature (in cm$^{-1}$).}
\label{Fig. 8}
\end{figure}

\begin{figure}[tpb]
%\begin{center}
\includegraphics[height=6cm,angle=0]{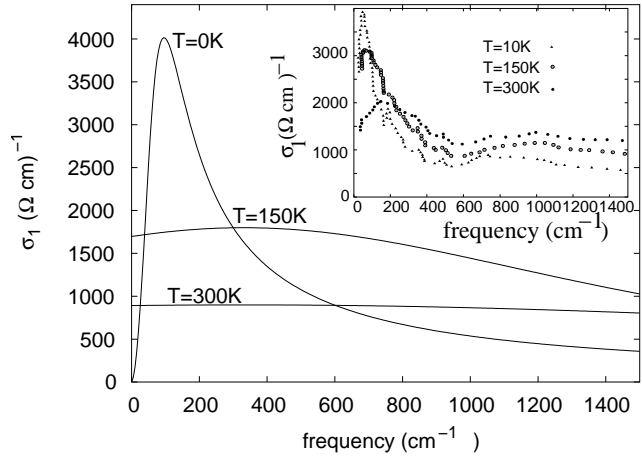}
\caption{Calculated AC conductivity as a function of frequency for
fixed hole density $\delta=0.184$ and different temperatures. {\sl
Inset:} AC conductivity versus frequency measured for a LSCO
sample with $x=0.184$ at different temperatures, taken from the
work of  Startseva {\it et al.}, Ref. \onlinecite{startseva}.}
\label{Fig. 9}
%\end{center}
\end{figure}

\begin{figure}[tbp]
\includegraphics[width=6cm,angle=0]{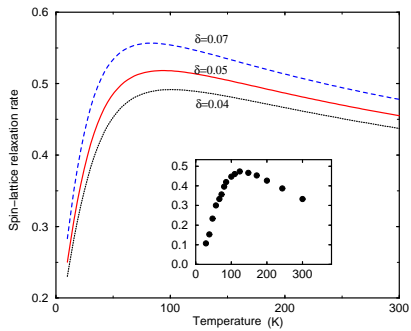}
\caption{ The calculated temperature dependence of spin-lattice relaxation rate $(1/T_1 T)^{63}$ for different dopings. Inset data of YBa$_2$Cu$_3$O$_{6.52}$ single crystal in units of $s^{-1}K^{-1}$, taken from Ref. \onlinecite{Ber}.}
\label{Fig. 10}
\end{figure}

\begin{figure}[tpb]
%\begin{center}
\includegraphics[height=6cm,angle=0]{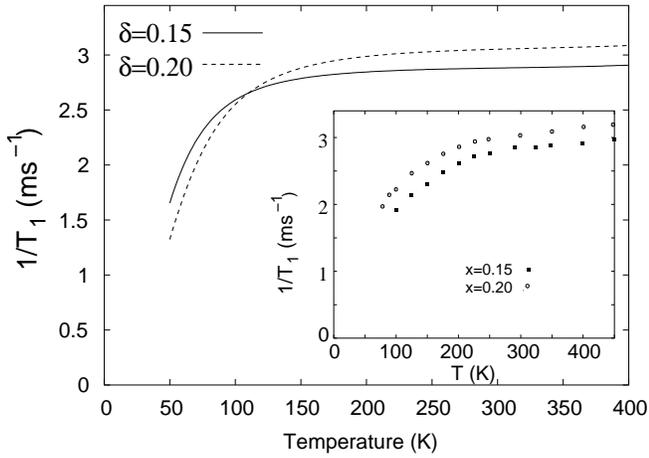}
\caption{Calculated temperature dependence of the spin-lattice
relaxation rate $1/T_1$ for different doping concentrations. {\sl
Inset:} spin-lattice relaxation rate measured in LSCO samples with
different Sr content $x$, taken from the work of Fujiyama {\it et
al.}, Ref.  \onlinecite{berthier}.} \label{Fig. 11}
%\end{center}
\end{figure}

\begin{figure}[tpb]
%\begin{center}
\includegraphics[width=6cm]{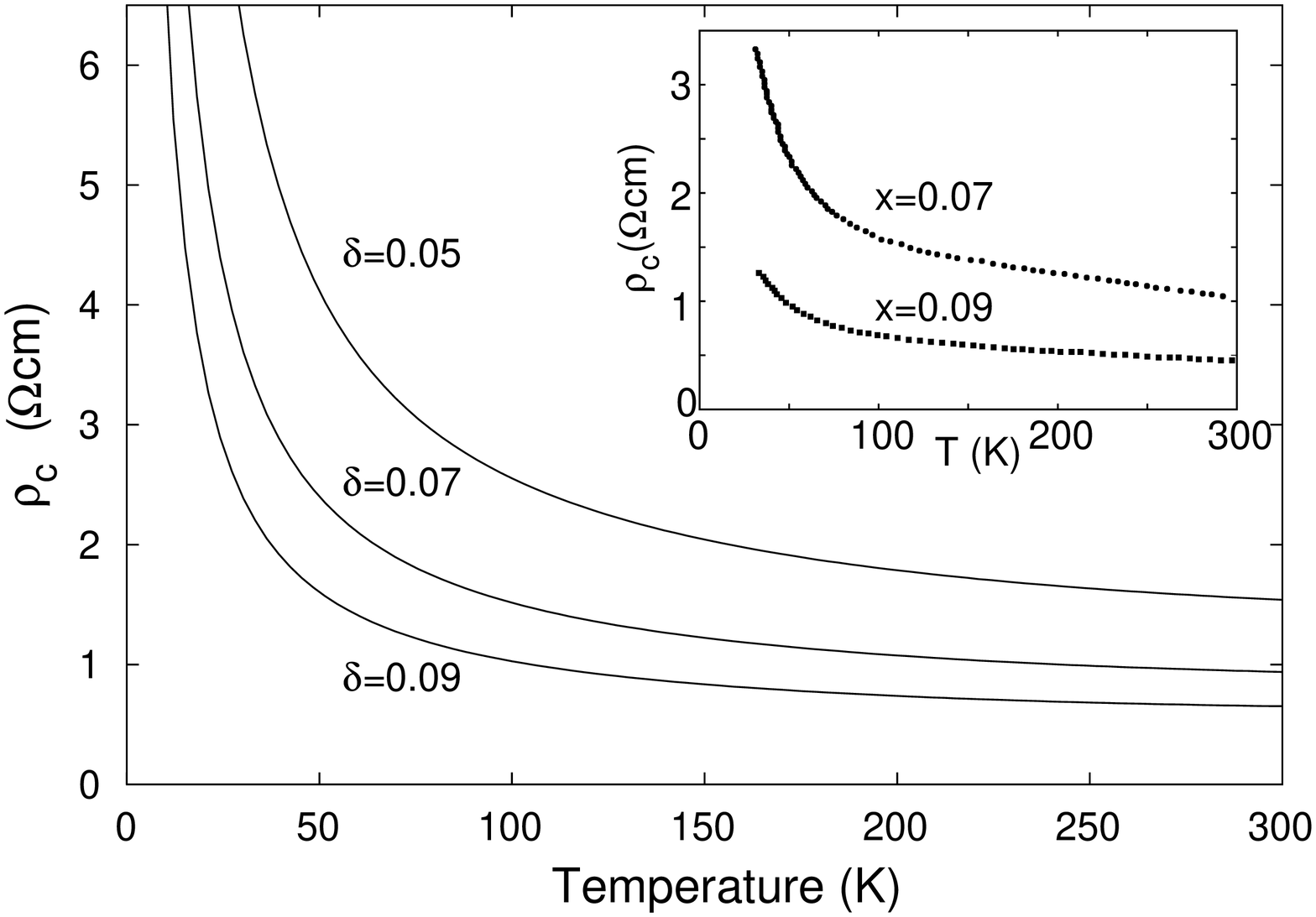}
\caption{ Calculated temperature dependence of the out-of-plane
resistivity (in arbitrary units) for different doping
concentrations: $\delta$=0.05 (full line), $\delta$=0.07 (dashed)
and $\delta$=0.09 (dotted). Inset shows experimental data on LSCO,
extracted from Ref. \onlinecite{magnes}}
 \label{Fig 12}
%\end{center}
\end{figure}

\begin{figure}[tpb]
%\begin{center}
\includegraphics[width=6cm]{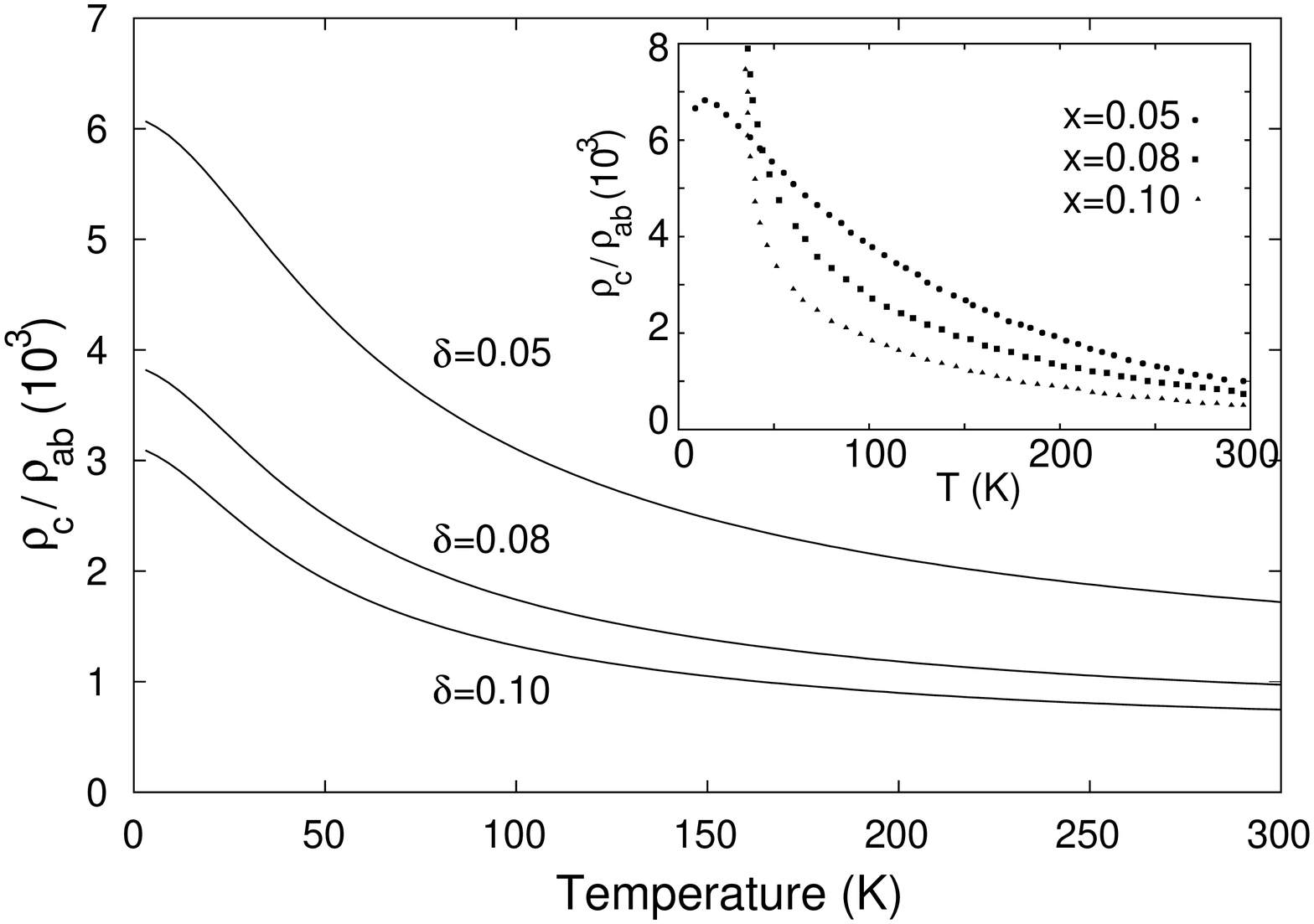}
\caption{ Calculated temperature dependence of the resistivity
anisotropy ratio as a function of
 temperature for different doping concentration: $\delta$=0.05 (full line),
 $\delta$ =0.07 (dashed) and  $\delta$=0.09 (dotted). Inset shows corresponding experimental data
 on LSCO, extracted from Ref. \onlinecite{Komiya}}
\label{Fig. 13}
%\end{center}
\end{figure}

\begin{figure}[tpb]
%\begin{center}
\includegraphics[height=6cm,angle=0]{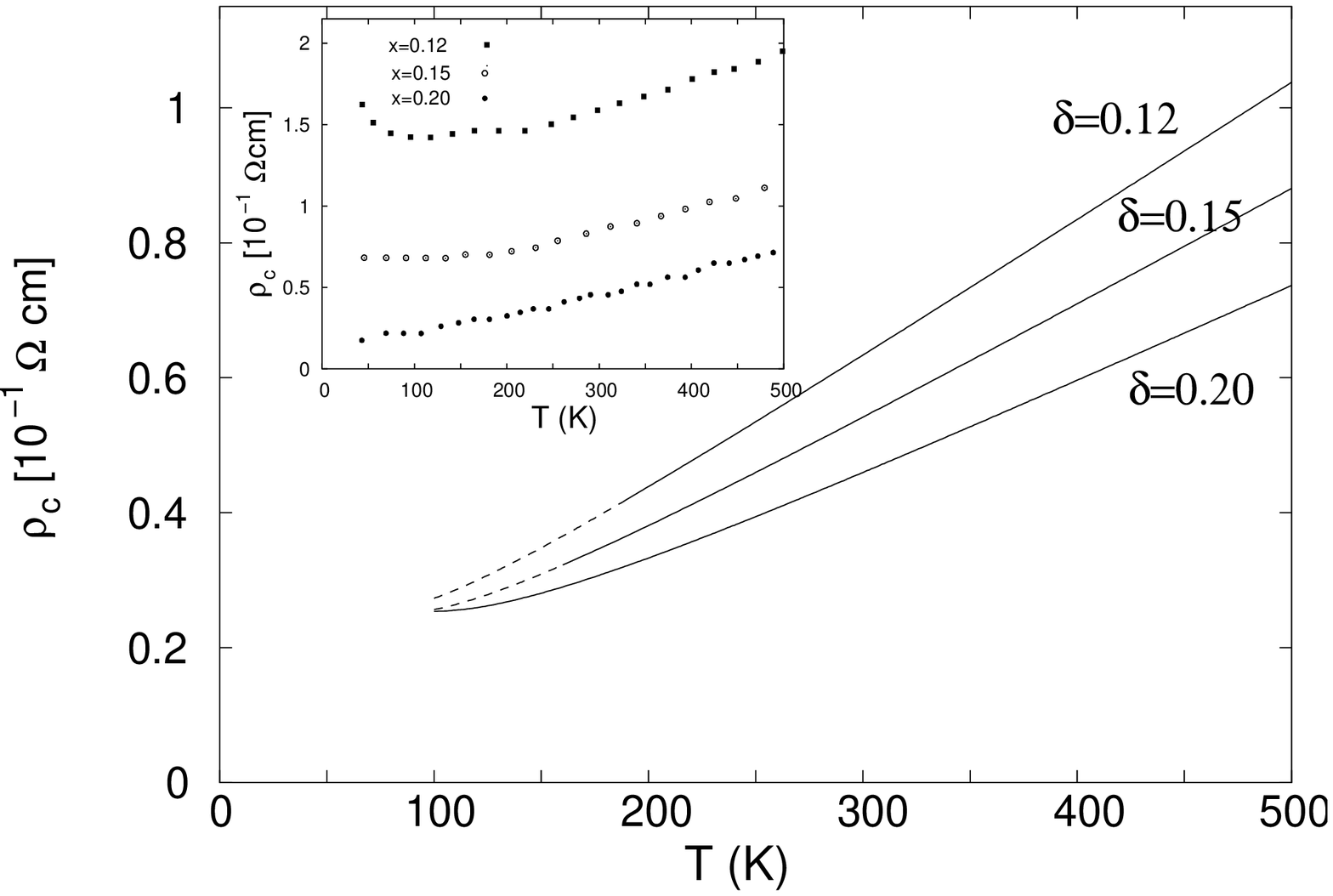}
\caption{Calculated out-of-plane resistivity as a function of
temperature for different hole concentrations. Below the
pseudo-gap temperature $T^*$, the curve is shown in dashed line.
{\sl Inset:} In-plane resistivity versus $T$ measured in LSCO
crystals with different Sr content $x$, taken from the work of
Nakamura {\it et al.}, Ref.  \onlinecite{nakamura}}.
\label{Fig. 14}
%\end{center}
\end{figure}

\begin{figure}[tpb]
%\begin{center}
\includegraphics[height=6cm,angle=0]{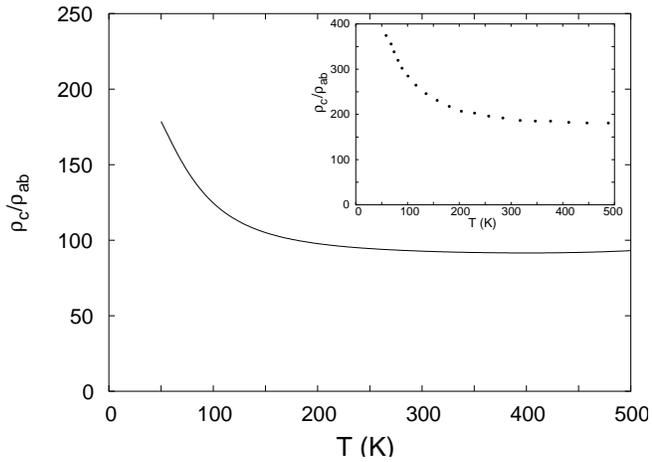}
\caption{Calculated resistivity anisotropy  ratio as a function of
temperature for fixed hole density $\delta=0.2$. {\sl Inset:}
resistivity anisotropy ratio measured for a LSCO sample with Sr
content $x=0.20$, taken from the work of Nakamura {\it et al.},
Ref.  \onlinecite{nakamura}.} \label{Fig. 15}
%\end{center}
\end{figure}

%\clearpage \clearpage

\end{document}